\documentclass[manuscript]{aastex}

\usepackage{amsbsy,amssymb,amsmath,bm}
\usepackage{graphicx,subfigure}
\usepackage{natbib}

\begin{document}

\title{Calculating rotating hydrodynamic and magneto-hydrodynamic waves to understand magnetic effects on dynamical tides}
\author{Xing Wei$^{1,2}$}
\affil{$^1$Institute of Natural Sciences and Department of Physics and Astronomy, Shanghai Jiao Tong University \\ $^2$Princeton University Observatory, Princeton, NJ 08544, USA}
\email{xing.wei@sjtu.edu.cn, xingwei@astro.princeton.edu}

\begin{abstract}
For understanding magnetic effects on dynamical tides, we study the rotating magneto-hydrodynamic (MHD) flow driven by harmonic forcing. The linear responses are analytically derived in a periodic box under the local WKB approximation. Both the kinetic and Ohmic dissipations at the resonant frequencies are calculated and the various parameters are investigated. Although magnetic pressure may be negligible compared to thermal pressure, magnetic field can be important for the first-order perturbation, e.g. dynamical tides. It is found that magnetic field splits the resonant frequency, namely the rotating hydrodynamic flow has only one resonant frequency but the rotating MHD flow has two, one positive and the other negative. In the weak field regime the dissipations are asymmetric around the two resonant frequencies and this asymmetry is more striking with a weaker magnetic field. It is also found that both the kinetic and Ohmic dissipations at the resonant frequencies are inversely proportional to the Ekman number and the square of wavenumber. The dissipation at the resonant frequency on small scales is almost equal to the dissipation at the non-resonant frequencies, namely the resonance takes its effect on the dissipation at intermediate length scales. Moreover, the waves with phase propagation perpendicular to magnetic field are much more damped. It is also interesting to find that the frequency-averaged dissipation is constant. This result suggests that in compact objects magnetic effects on tidal dissipation should be considered.

\end{abstract}

\maketitle

\section{Introduction}

Tides exist widely in binary systems, e.g. Earth-Moon, binary stars, exoplanet and host star, etc. The tidal torque transfers angular momentum between the orbital motion and the rotational motion of the binary components such that the orbital and rotational frequencies eventually become equal (synchronization) and the orbit eventually becomes circular (circularization). In the process of synchronisation and circularisation, the dissipation in the fluid interior of the star or planet plays an important role. There are two parts of the response to the tidal force, the wave part and the non-wave part. The non-wave part is a large-scale deformation in the quasi-hydrostatic balance, called the equilibrium tide. The wave part is the fluid waves excited by the tidal force, called the dynamical tide. Because the dynamical tide has a much smaller length scale than the equilibrium tide, it can be more important for the tidal dissipation. When the eigen-frequencies of these waves are close to the frequency of the tidal force, resonance occurs, at which the response and the dissipation are greatly increased.

For the dynamical tide, sound waves and surface gravity waves have frequencies too large to be resonantly excited, but internal gravity waves due to density stratification and inertial waves due to rotation can be excited. When the tidal frequency is close to the buoyancy frequency, internal gravity waves are excited in the stably stratified region (e.g. the radiation zone). When the tidal frequency is close to the rotational frequency, inertial waves are excited. The problem of the dynamical tide in stellar interiors was firstly considered by \citet{cowling}. Internal gravity waves were studied by \citet{zahn} and then applied to the interpretation of the angular momentum transfer in the stellar radiation zone by \citet{goldreich} and \citet{goodman1998}. Later, internal gravity waves due to a compositional jump were studied by \citet{lai}. The problem of the inertial waves in spherical geometry is more difficult, because, firstly, rotation breaks the symmetry of the equation of fluid motion such that the radial and colatitude directions are coupled (for comparison, the equation of the internal gravity waves can be reduced to a one-dimensional eigenvalue problem in the radial direction), and secondly, the governing equation (Poincar\'e equation) of the inviscid inertial waves is singular at the critical latitude and viscosity smooths singularity such that the inertial waves are spawned at the critical latitude and propagate in the thin shear layers because of wave reflection, i.e. the wave attractors \citep{busse,hollerbach1,rieutord,ogilvie2005,tilgner,zhang}. Recently, tidally excited inertial waves were studied both analytically and numerically by \citet{ogilvie2004,wu1,wu2,goodman2009,ogilvie2014}, etc. Studies of dynamical tides were summarized in the review paper by \citet{ogilvie-review}.

However, magnetic effects on dynamical tides have not been extensively studied. \citet{kerswell} once studied the MHD waves excited by the tide in the Earth's core and focused on the elliptical instability, and the magneto-elliptic-rotational waves were also studied in \citet{goodman1993,mizerski-bajer2009,mizerski-bajer2011,mizerski-bajer-moffatt2012}, but there have been few studies on magnetic dynamical tides, i.e. the {\bf magneto-inertial waves}. Although the magnetic field is not strong on the stellar surface, it might be strong in the stellar interior because of the dynamo action. Moreover, even if the magnetic field is insignificant for the equilibrium state (in the sense that the magnetic pressure is small compared to the thermal pressure), it can be important for the dynamics of, say, the first-order perturbation. As is known, magnetically modified inertial waves (i.e. the magneto-inertial wave) have very different frequencies from non-magnetic inertial waves. In addition, inertial or magneto-inertial waves with helical spatial structure can support dynamo action through the $\alpha$ effect to reinforce the magnetic field \citep{moffatt-kinematic,moffatt-dynamic,wei}. Therefore, the magnetic field and the waves are mutually interacting.

In this paper we will study the magnetically modified inertial waves (i.e. magneto-inertial waves) excited by the tidal forcing and focus on resonances. We will use the simplified geometry of a periodic box to perform our study. The purpose is to understand how the magnetic field influences the resonant frequency and hence the tidal dissipation. Both the kinetic dissipation and the Ohmic dissipation will be studied. In Section \S\ref{sec:response} the linear response to the tidal forcing in the rotating MHD flow is derived and the resonant frequencies are given. In Section \S\ref{sec:dissipation} the explicit expressions for the calculation of the tidal dissipation are given in the dimensionless form. In Section \S\ref{sec:hydro} the results of the rotating hydrodynamic flow in the absence of magnetic field are shown. In Section \S\ref{sec:mhd} the results of the rotating MHD flow in the presence of magnetic field are shown. In Section \$\ref{sec:application} some astrophysical applications are discussed. In Section \S\ref{sec:discussion} a brief summary and some further discussions are given.

\section{Linear response and resonant frequency}\label{sec:response}

Because the frequencies of sound waves are too high to be resonantly excited by the dynamical tide, we study the incompressible fluid. The derivation of the unforced rotating MHD system can be found in \S10.2 in \citet{moffatt-book}. We extend this derivation to the forced system. The Navier-Stokes equation of the incompressible MHD in the rotating frame at the constant angular velocity $\bm\Omega$ reads
\begin{equation}
\frac{\partial\bm u}{\partial t}+\bm u\cdot\bm\nabla\bm u=-\frac{1}{\rho}\bm\nabla p+\nu\nabla^2\bm u+2\bm u\times\bm\Omega+\frac{1}{\rho\mu}\bm B\cdot\bm\nabla\bm B,
\end{equation}
where the induced pressure $p$ includes the centrifugal force $|\bm\Omega\times\bm x|^2/2$ and the magnetic pressure $B^2/(2\mu)$. The continuity equation of the incompressible fluid reads
\begin{equation}
\bm\nabla\cdot\bm u=0.
\end{equation}
The magnetic induction equation reads
\begin{equation}
\frac{\partial\bm B}{\partial t}+\bm u\cdot\bm\nabla\bm B=\bm B\cdot\bm\nabla\bm u+\eta\nabla^2\bm B.
\end{equation}
The solenoidal condition of magnetic field is
\begin{equation}
\bm\nabla\cdot\bm B=0.
\end{equation}
We then assume that the length scales of spatial variation of the background flow $\bm u_0$ (i.e. the mean flow) and the background field $\bm B_0$ are much larger than those of perturbations such that $\bm u_0$ and $\bm B_0$ can be considered to be uniform. This is the local WKB approximation. Suppose that 
\begin{equation}\label{eq:decompose}
\bm u=\bm u_0+\bm u_1, \hspace{3mm} p=p_0+p_1, \hspace{3mm} \bm B=\bm B_0+\bm B_1,
\end{equation}
where $\bm u_1$, $p_1$ and $\bm B_1$ are the first-order Eulerian perturbations. The substitution of \eqref{eq:decompose} into the Navier-Stokes and magnetic induction equations with the neglect of quadratic terms yields 
\begin{equation}\label{eq:ns}
\frac{\partial\bm u_1}{\partial t}+\bm u_0\cdot\bm\nabla\bm u_1=-\frac{1}{\rho}\bm\nabla p_1+\nu\nabla^2\bm u_1+2\bm u_1\times\bm\Omega+\frac{1}{\rho\mu}\bm B_0\cdot\bm\nabla\bm B_1+\bm f,
\end{equation}
where $\bm f$ is the force to excite waves and corresponds to the tidal force. It should be noted that the external force appears only in the perturbation equation. The perturbed magnetic induction equation reads
\begin{equation}\label{eq:induction}
\frac{\partial\bm B_1}{\partial t}+\bm u_0\cdot\bm\nabla\bm B_1=\bm B_0\cdot\bm\nabla\bm u_1+\eta\nabla^2\bm B_1.
\end{equation}

We then take out a small piece of region in the stellar or planetary interior. The size of this region is small compared to the length scale of the background flow and field such that this region can be considered to be subject to the periodic boundary condition. Moreover, we assume that the driving force is a single traveling wave on top of the background flow, namely
\begin{equation}\label{eq:force}
\bm f=\Re\{\hat{\bm f}\exp\left[i\left(\bm k\cdot\bm x-(\omega+\bm u_0\cdot\bm k)t\right)\right]\},
\end{equation}
where $\hat{\bm f}$ is the complex amplitude, $\bm k$ the wavevector, $\omega$ the frequency and $\Re$ denotes taking the real part. Because it is a linear problem, equations \eqref{eq:ns} and \eqref{eq:induction} admit the solution of the form
\begin{equation}\label{eq:solution}
(\bm u_1, p_1, \bm B_1)=\Re\{(\hat{\bm u}, \hat p, \hat{\bm B})\exp\left[i\left(\bm k\cdot\bm x-(\omega+\bm u_0\cdot\bm k)t\right)\right]\}.
\end{equation}
Substituting \eqref{eq:solution} into \eqref{eq:induction}, we derive
\begin{equation}\label{eq:B-u}
\hat{\bm B}=-\frac{\bm k\cdot\bm B_0}{\omega+i\eta k^2}\hat{\bm u}.
\end{equation}
Substituting \eqref{eq:solution} into \eqref{eq:ns} and using \eqref{eq:B-u}, we are led to
\begin{equation}\label{eq:u}
-i\sigma\hat{\bm u}+2\bm\Omega\times\hat{\bm u}=-i\bm k\frac{\hat p}{\rho}+\hat{\bm f},
\end{equation}
where
\begin{equation}
\sigma=(\omega+i\nu k^2)-\frac{|\bm k\cdot\bm B_0|^2}{\rho\mu(\omega+i\eta k^2)}.
\end{equation}
The background magnetic field, viscosity and magnetic diffusivity are entirely contained in the coefficient $\sigma$. We can express $\sigma$ in the simpler form,
\begin{equation}\label{eq:sigma}
\sigma=(\omega+i\nu k^2)-\frac{\omega_B^2}{\omega+i\eta k^2},
\end{equation}
where 
\begin{equation}
\omega_B=\frac{\bm k\cdot\bm B_0}{\sqrt{\rho\mu}}
\end{equation}
is the frequency of Alfv\'en wave. In the absence of magnetic field, $\omega_B=0$, the problem reduces to the tidal resonance of inertial wave in the rotating hydrodynamic flow.

To give the driving force $\bm f$, we come back to the tidal force for which $\bm f$ models. The tidal force is the difference between the force exerted by the perturbing body on any point in the interior of the primary body and the force exerted by the perturbing body at the centre of the primary body. It can be derived from the tidal potential, i.e. the superposition of spherical harmonics with harmonic dependence on time in terms of the Doppler-shifted frequency \citep{ogilvie-review}. So the tidal force is curl-free. However, its contribution to the dynamical tide is vortical because of the slow equilibrium tide, see the details in Appendix B of \citet{ogilvie2005}. Briefly speaking, the incompressible equilibrium tide varies slowly and does not satisfy the hydrostatic balance such that the residual is a vortical force that can drive the dynamical tide, e.g. the inertial waves in rotating fluid. In our model, $\bm f$ corresponds to the force responsible for the dynamical tide and it is not curl-free but vortical. On the other hand, to have the dynamical effect on an incompressible flow, the driving force $\bm f$ cannot be curl-free (if it is curl-free then it will be absorbed into the pressure term and act as the additional pressure). Although any vortical force can act as the driving force $\bm f$, we assume the driving force to be helical, i.e. vorticity is parallel to velocity such that the vortical effect reaches the maximum, namely helicity (the dot product of velocity and vorticity) reaches the maximum. One may argue that the helical force is too artificial. Here we give two reasons. Firstly, this assumption is for simplicity to derive the solution, see the next derivations, and this simplicity does not make physics of the tidal problem lost. Secondly, any vector field can be decomposed into the curl-free part and the divergence-free part, i.e. the Helmholtz decomposition. The divergence-free part can be further decomposed into helical modes, see \citet{waleffe}. Back to the driving force $\bm f$, the curl-free part can be absorbed into the pressure gradient and the divergence-free part can be expressed as the superposition of helical forces. For a linear problem, we study the tidal wave excited by a single helical force. This is the reason that we use the helical force for the study of tidal waves. Consequently, $\bm f$ satisfies
\begin{equation}\label{eq:helical}
i\bm k\times\hat{\bm f}=k\hat{\bm f}.
\end{equation}

Performing $\bm k\times$ on \eqref{eq:u} to eliminate pressure and using $\bm k\cdot\hat{\bm u}=0$ (incompressible fluid) and \eqref{eq:helical}, we derive
\begin{equation}\label{eq:curl}
-i\sigma\bm k\times\hat{\bm u}-(2\bm k\cdot\bm\Omega)\hat{\bm u}=-ik\hat{\bm f}.
\end{equation}
Performing $\bm k\times$ again on \eqref{eq:curl}, we derive
\begin{equation}\label{eq:double-curl}
i\sigma k^2\hat{\bm u}-(2\bm k\cdot\bm\Omega)\bm k\times\hat{\bm u}=-k^2\hat{\bm f}.
\end{equation}
Combining \eqref{eq:curl} and \eqref{eq:double-curl} to eliminate $\bm k\times\hat{\bm u}$ leads to
\begin{equation}
(2\bm k\cdot\bm\Omega+k\sigma)(2\bm k\cdot\bm\Omega-k\sigma)\hat{\bm u}=ik(2\bm k\cdot\bm\Omega-k\sigma)\hat{\bm f}.
\end{equation}
According to \eqref{eq:sigma}, in the presence of small viscosity $\nu$ or magnetic diffusivity $\eta$ in the real geophysical and astrophysical fluids, $\sigma$ cannot be a real number such that the non-zero factor $(2\bm k\cdot\bm\Omega-k\sigma)$ can be cancelled, and thus we are led to
\begin{equation}
\hat{\bm u}=\frac{i\hat{\bm f}}{\sigma+2\bm k\cdot\bm\Omega/k}.
\end{equation}
We can express the solution in the simpler form
\begin{equation}\label{eq:resonance}
\hat{\bm u}=\frac{i\hat{\bm f}}{\sigma+\omega_\Omega},
\end{equation}
where 
\begin{equation}
\omega_\Omega=\frac{2\bm k\cdot\bm\Omega}{k}
\end{equation}
is the frequency of inertial wave. Equation \eqref{eq:resonance} is the solution of the linear response, where $\sigma$ is given by \eqref{eq:sigma}.

In equation \eqref{eq:resonance} the singularity cannot occur, i.e. $\sigma\neq-\omega_\Omega$, due to the presence of viscosity or magnetic diffusivity (see equation \eqref{eq:sigma}). However, the response $\hat{\bm u}$ becomes very strong at some particular forcing frequencies when the condition $\sigma=-\omega_\Omega$ is satisfied with both viscosity and magnetic diffusivity neglected. This situation is called the {\bf resonance}. Accordingly, the frequency $\omega$ given by 
\begin{equation}\label{eq:resonant-frequency}
\omega-\frac{\omega_B^2}{\omega}=-\omega_\Omega
\end{equation}
is the {\bf resonant frequency}. In the rotating hydrodynamic flow, magnetic field is absent and equation \eqref{eq:resonant-frequency} yields only one resonant frequency, i.e. the inertial wave
\begin{equation}\label{eq:resonant-frequency-rotating}
\omega_0=-\omega_\Omega=-\frac{2\bm k\cdot\bm\Omega}{k}.
\end{equation}
In the rotating MHD flow, the quadratic equation \eqref{eq:resonant-frequency} yields two resonance frequencies, i.e. the magneto-inertial waves,
\begin{equation}\label{eq:resonant-frequency-solution}
\omega^{\pm}=\frac{1}{2}\left(-\omega_\Omega\pm\sqrt{\omega_\Omega^2+4\omega_B^2}\right)=-\frac{\bm k\cdot\bm\Omega}{k}\pm\sqrt{\left(\frac{\bm k\cdot\bm\Omega}{k}\right)^2+\frac{|\bm k\cdot\bm B_0|^2}{\rho\mu}}.
\end{equation}
In the case of $\bm k\cdot\bm\Omega>0$, the positive solution $\omega^+$ and the negative solution $\omega^-$ satisfy
\begin{equation}
\omega^-<\omega_0<0<\omega^+.
\end{equation}
In the case of $\bm k\cdot\bm\Omega<0$, the sorting becomes
\begin{equation}
\omega^-<0<\omega_0<\omega^+.
\end{equation}
In both the cases, it follows that
\begin{equation}
\omega^-<\omega_0<\omega^+.
\end{equation}
It indicates that the presence of magnetic field broadens the range of the resonant frequency of rotating hydrodynamic flow, such that the tidal resonance with magnetic field can occur more possibly. 

A special case of magneto-inertial wave is the magnetostrophic wave with the neglect of $\partial\bm u/\partial t$ in the perturbed Navier-Stokes equation, i.e. the magnetostrophic balance of pressure, Corilis force and Lorentz force. This magnetostrophic wave is slow and long, and may contribute to the geodynamo in the Earth's fluid core. The readers who are interested can find the details about this wave in \citet{moffatt-book,schmitt,davidson,wei_thesis}.

\section{Dissipation, driving force and normalisation}\label{sec:dissipation}

For astronomy and astrophysics, tidal dissipation is paid more attention than tidal response, because the former determines the orbital evolution of binary system. We now calculate the dissipation. With the periodic boundary condition, the volume-averaged kinetic dissipation $D_k$ can be calculated as
\begin{equation}\label{eq:kinetic-dissipation}
D_k=\frac{1}{V}\int\rho\nu|\bm\nabla\times\bm u_1|^2dV=\frac{\rho\nu}{2}|i\bm k\times\hat{\bm u}|^2,
\end{equation}
where $\hat{\bm u}$ is given by \eqref{eq:resonance}. In the MHD flow, in addition to the kinetic dissipation, the Ohmic dissipation is important and it can be calculated as
\begin{equation}\label{eq:ohmic-dissipation}
D_m=\frac{1}{V}\int_V\frac{\eta}{\mu}|\bm\nabla\times\bm B_1|^2dV=\frac{\eta}{2\mu}|i\bm k\times\hat{\bm B}|^2=\frac{\rho\eta}{2}\frac{\omega_B^2}{|\omega+i\eta k^2|^2}|i\bm k\times\hat{\bm u}|^2.
\end{equation}
In the above derivation equation \eqref{eq:B-u} is used.

To explicitly calculate the dissipations, we use the Cartesian coordinate system ($x,y,z$) and the small piece of region is considered as a periodic cube with its size being $l$. Thus, the wavevector is given to be
\begin{equation}\label{eq:wavevector}
\bm k=(k_x,k_y,k_z)=(2\pi n_x/l,2\pi n_y/l, 2\pi n_z/l),
\end{equation}
where $n_x$, $n_y$ and $n_z$ are integers for periodicity. In the local coordinate system, we choose the $z$ axis along the angular velocity $\bm\Omega$ and the plane of $\bm\Omega$ and $\bm B_0$ to be the $x-z$ plane (if $\bm B_0$ is parallel or anti-parallel to $\bm\Omega$ then the $x$ axis is arbitrary as long as it is perpendicular to the $z$ axis). Therefore, $\bm B_0$ is expressed as
\begin{equation}\label{eq:B0}
\bm B_0=(B_0\sin\alpha,0,B_0\cos\alpha),
\end{equation}
where $\alpha$ is the angle between $\bm\Omega$ and $\bm B_0$.

We also need to find the explicit expression of $\hat{\bm f}$. Equation \eqref{eq:helical} is degenerate (i.e. only two components are independent) and yields
\begin{equation}\label{eq:degenerate}
\frac{\hat f_y}{\hat f_x}=\frac{-k_xk_y+ikk_z}{k_y^2+k_z^2}, \;
\frac{\hat f_z}{\hat f_y}=\frac{-k_yk_z+ikk_x}{k_z^2+k_x^2}, \;
\frac{\hat f_x}{\hat f_z}=\frac{-k_zk_x+ikk_y}{k_x^2+k_y^2}.
\end{equation}
We denote the force amplitude by $a$, i.e. 
\begin{equation}\label{eq:amplitude=a}
|\hat{\bm f}|=\sqrt{|\hat f_x|^2+|\hat f_y|^2+|\hat f_z|^2}=\sqrt{\hat f_x\hat f_x^*+\hat f_y\hat f_y^*+\hat f_z\hat f_z^*}=a,
\end{equation}
where $^*$ denotes the complex conjugate. Equations \eqref{eq:degenerate} and \eqref{eq:amplitude=a} then combine to yield
\begin{equation}\label{eq:amplitude}
|\hat f_x|=\frac{\sqrt{k_y^2+k_z^2}}{\sqrt{2}k}a, \; 
|\hat f_y|=\frac{\sqrt{k_z^2+k_x^2}}{\sqrt{2}k}a, \; 
|\hat f_z|=\frac{\sqrt{k_x^2+k_y^2}}{\sqrt{2}k}a,
\end{equation}
and in addition, the arguments of $\hat f_y/\hat f_x$ and $\hat f_z/\hat f_x$ are, respectively,
\begin{equation}\label{eq:argument}
\pi-\arccos\frac{k_xk_y}{\sqrt{(k_y^2+k_z^2)(k_z^2+k_x^2)}} \hspace{3mm} \mbox{and} \hspace{3mm}
\pi+\arccos\frac{k_zk_x}{\sqrt{(k_x^2+k_y^2)(k_y^2+k_z^2)}}.
\end{equation}
The arguments of $\hat f_x$, $\hat f_y$ and $\hat f_z$ themselves are insignificant for the volume integral of energy and dissipation, but the differences between them do matter, and so the argument of $\hat f_x$ is given to be $0$. Thus, equations \eqref{eq:amplitude} and \eqref{eq:argument} give the three components of the complex amplitude $\hat{\bm f}$.

Usually the dimensionless calculation is preferred because it is more physically meaningful. We normalise length with $l$, time with $\Omega^{-1}$, velocity with $l\Omega$, force amplitude with $l\Omega^2$, magnetic field with $B_0$ and the two dissipations $D_k$ and $D_m$ with $\rho l^2\Omega^3$. For simplicity we use the notation of the dimensional quantities for the dimensionless quantities, but we need to keep in mind that from now on all the physical variables are dimensionless. The dimensionless version for the calculation of $D_k$ and $D_m$ is then translated to be
\begin{equation}\label{eq:kinetic-dissipation-dimensionless}
D_k=\frac{E}{2}|i\bm k\times\hat{\bm u}|^2,
\end{equation}
\begin{equation}\label{eq:ohmic-dissipation-dimensionless}
D_m=\frac{E}{2Pm}Le^2\frac{(k_x\sin\alpha+k_z\cos\alpha)^2}{|\omega+i\frac{E}{Pm}k^2|^2}|i\bm k\times\hat{\bm u}|^2,
\end{equation}
\begin{equation}\label{eq:resonance-dimensionless}
\hat{\bm u}=\frac{i\hat{\bm f}}{\sigma+2k_z/k},
\end{equation}
\begin{equation}\label{eq:sigma-dimensionless}
\sigma=(\omega+iEk^2)-Le^2\frac{(k_x\sin\alpha+k_z\cos\alpha)^2}{\omega+i\frac{E}{Pm}k^2}.
\end{equation}
In the above dimensionless equations, the Ekman number
\begin{equation}\label{eq:ekman}
E=\frac{\nu}{l^2\Omega}
\end{equation}
measures the ratio of the rotational time scale to the viscous time scale, which is very small in the stellar and planetary interiors ($E\ll1$), the Lehnert number
\begin{equation}\label{eq:lehnert}
Le=\frac{B_0}{\sqrt{\rho\mu}l\Omega}
\end{equation}
measures the ratio of the rotational time scale to the Alfv\'enic time scale, and the magnetic Prandtl number 
\begin{equation}\label{eq:prandtl}
Pm=\frac{\nu}{\eta}
\end{equation}
measures the ratio of viscosity to magnetic diffusivity. The dimensionless resonant frequency of the rotating hydrodynamic flow is
\begin{equation}
\omega_0=-\frac{2k_z}{k}.
\end{equation}
The two resonant frequencies of the rotating MHD flow are
\begin{equation}\label{eq:resonant-frequency-solutions-dimensionless}
\omega^{\pm}=-\frac{k_z}{k}\pm\sqrt{\left(\frac{k_z}{k}\right)^2+Le^2(k_x\sin\alpha+k_z\cos\alpha)^2}.
\end{equation}

Because it is a linear problem, the dissipation scales as the square of the force amplitude, i.e. $D\propto a^2$. Thus we fix $a=1$ in this paper. In the next two sections we will calculate $D_k$ \eqref{eq:kinetic-dissipation-dimensionless} and $D_m$ \eqref{eq:ohmic-dissipation-dimensionless} according to \eqref{eq:resonance-dimensionless} and \eqref{eq:sigma-dimensionless}.

\section{Results of the rotating hydrodynamic flow}\label{sec:hydro}

\begin{figure}
\centering
\includegraphics[scale=0.4]{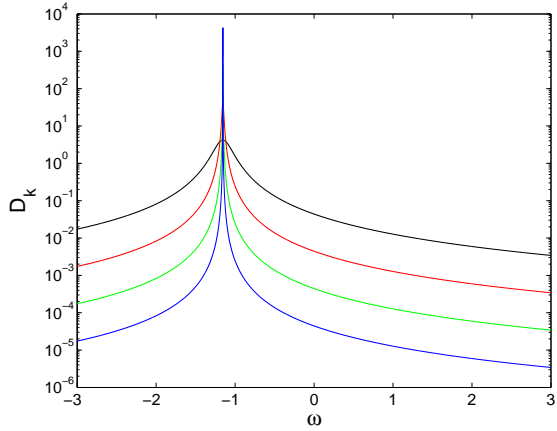}
\caption{The rotating hydrodynamic flow. The kinetic dissipation $D_k$ versus the force frequency $\omega$. The black, red, green and blue lines denote respectively $E=10^{-3}$, $10^{-4}$, $10^{-5}$ and $10^{-6}$. $n_x=n_y=n_z=1$.}\label{fig:hydro1}
\end{figure}

Before studying the rotating MHD flow, we study the rotating hydrodynamic flow in the absence of magnetic field, i.e. $Le=0$. In this case, the resonance occurs at 
\begin{equation}
\omega_0=-\frac{2k_z}{k}. 
\end{equation}
Therefore the orientation of the force wavevector determines the resonant frequency and the dissipation. We try firstly the lowest wavenumbers $n_x=n_y=n_z=1$ and so the resonant frequency is $\omega_0=-2/\sqrt{3}\approx-1.1547$. Figure \ref{fig:hydro1} shows the kinetic dissipation $D_k$ versus the force frequency $\omega$. It reveals that the dissipation becomes very strong at the resonant frequency $-1.1547$. It also suggests that the dissipation at the resonant frequency is stronger at the lower $E$ but the dissipation at the other frequencies is weaker at the lower $E$.

By virtue of \eqref{eq:resonance-dimensionless} and \eqref{eq:sigma-dimensionless} we can explicitly derive the velocity at the resonant frequency $\omega_0=-2k_z/k$ to be
\begin{equation}\label{eq:u-f}
\hat{\bm u}=\frac{1}{Ek^2}\hat{\bm f}.
\end{equation}
This indicates that the kinetic dissipation $D_k\propto Ek^2|\hat{\bm u}|^2$ at the resonant frequency scales as
\begin{equation}\label{eq:scaling}
D_k\propto E^{-1}k^{-2}.
\end{equation}

\begin{figure}
\centering
\includegraphics[scale=0.4]{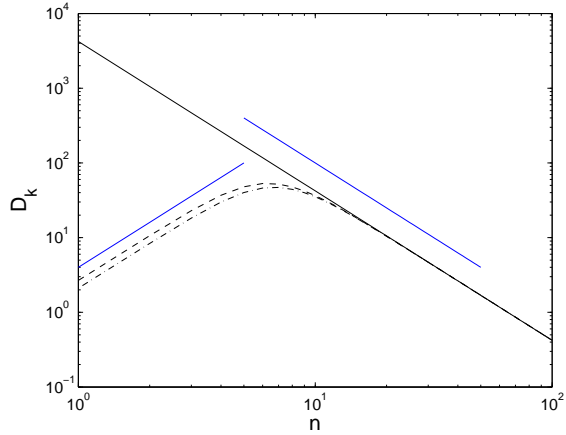}
\caption{The rotating hydrodynamic flow. The kinetic dissipation $D_k$ versus the force wavenumbers $n_x=n_y=n_z=n$. The solid line denotes the resonant frequency $\omega_0=-2/\sqrt{3}\approx-1.1547$, the dashed line $\omega=-1.15$ and the dash-dotted line $\omega=-1.16$. The two blue straight lines show the two scalings $n^{-2}$ and $n^2$. $E=10^{-6}$.}\label{fig:hydro2}
\end{figure}

We then study the effect of the force wavenumber on the dissipation. We keep the orientation of the force wavevector, i.e. $n_x=n_y=n_z$, but increases its magnitude such that the resonant frequency is always $\omega_0=-2/\sqrt{3}\approx-1.1547$ (changing its orientation simply shifts the resonant frequency). Figure \ref{fig:hydro2} shows $D_k$ versus wavenumbers $n_x=n_y=n_z=n$. In addition to the dissipation at the resonant frequency we also calculate the dissipation at the other two frequencies $-1.15$ and $-1.16$ neighbouring to the resonant frequency (one greater and the other less than the resonant frequency). It verifies that $D_k\propto k^{-2}$ at the resonant frequency. Moreover, the dissipation at the other frequencies scales as $k^2$ for the low wavenumbers and becomes equal to the dissipation at the resonant frequency for the high wavenumbers (in this case for the wavenumbers higher than $20$). In this sense, the resonance takes its effect on the dissipation at intermediate length scales (large compared to the box size but still much smaller than the length scale of the background flow and field). The wavenumber at which the non-resonant dissipation reaches its peak can be roughly estimated by $k_{\rm peak}\approx\sqrt{|\omega-\omega_0|/E}$, which is derived by equating the derivative of $D_k$ to zero (see \eqref{eq:kinetic-dissipation-dimensionless}, \eqref{eq:resonance-dimensionless}, \eqref{eq:sigma-dimensionless}).

\begin{figure}
\centering
\includegraphics[scale=0.4]{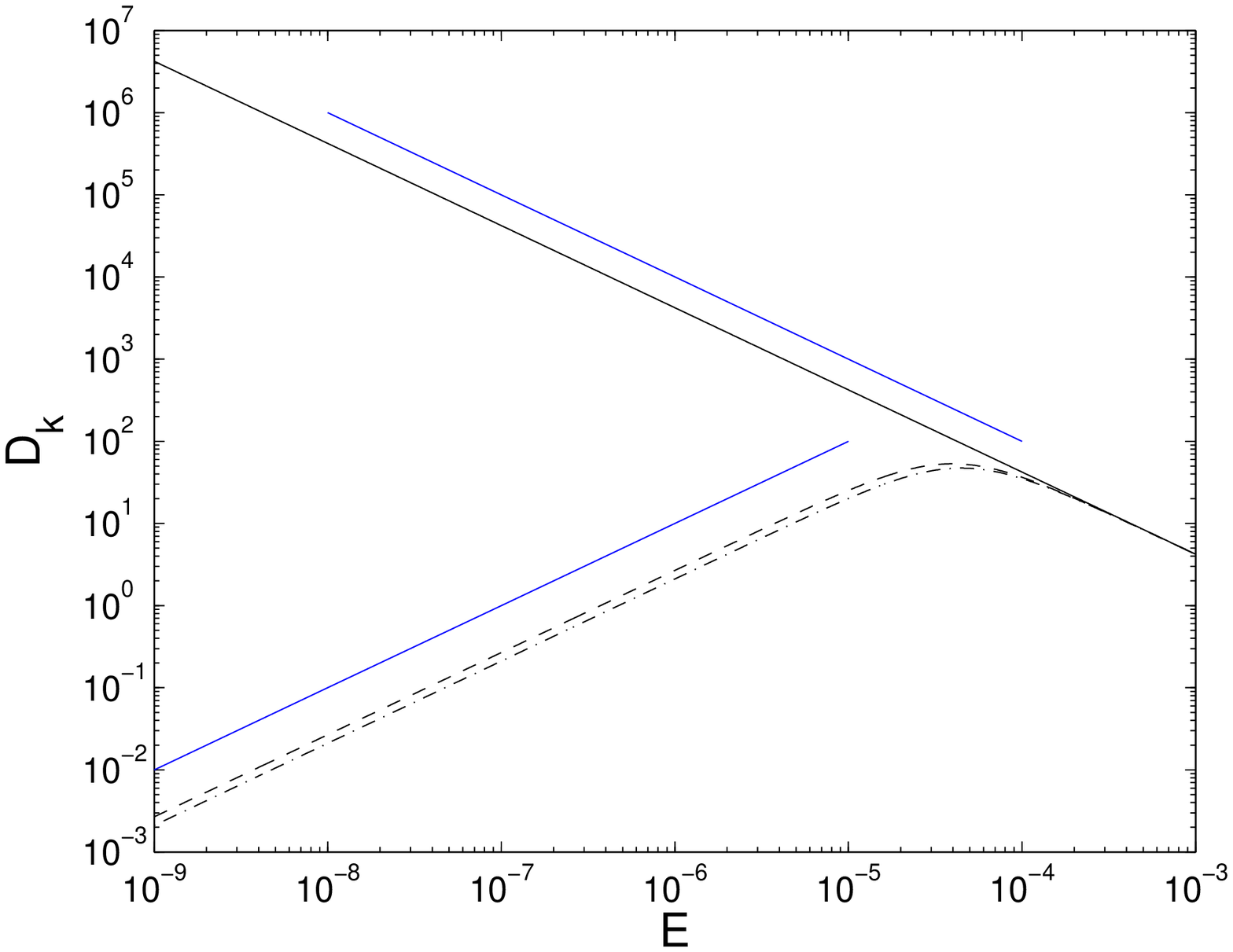}
\caption{The rotating hydrodynamic flow. The kinetic dissipation $D_k$ versus the Ekman number $E$. The solid line denotes the resonant frequency $\omega_0=-2/\sqrt{3}\approx-1.1547$, the dashed line $\omega=-1.15$ and the dash-dotted line $\omega=-1.16$. The two blue straight lines show the two scalings $E^{-1}$ and $E$. $n_x=n_y=n_z=1$.}\label{fig:hydro3}
\end{figure}

Equation \eqref{eq:scaling} also indicates that the kinetic dissipation $D_k$ at the resonant frequency scales as $E^{-1}$. Figure \ref{fig:hydro3} shows that the kinetic dissipation at the resonant frequency scales as $E^{-1}$, and at the other frequencies the kinetic dissipation scales as $E$ at the low $E$ and reaches the level of the resonant frequency at the high $E$. In this sense, the resonance takes its effect on the dissipation on small Ekman number.

\section{Results of the rotating MHD flow}\label{sec:mhd}

\begin{figure}
\centering
\includegraphics[scale=0.4]{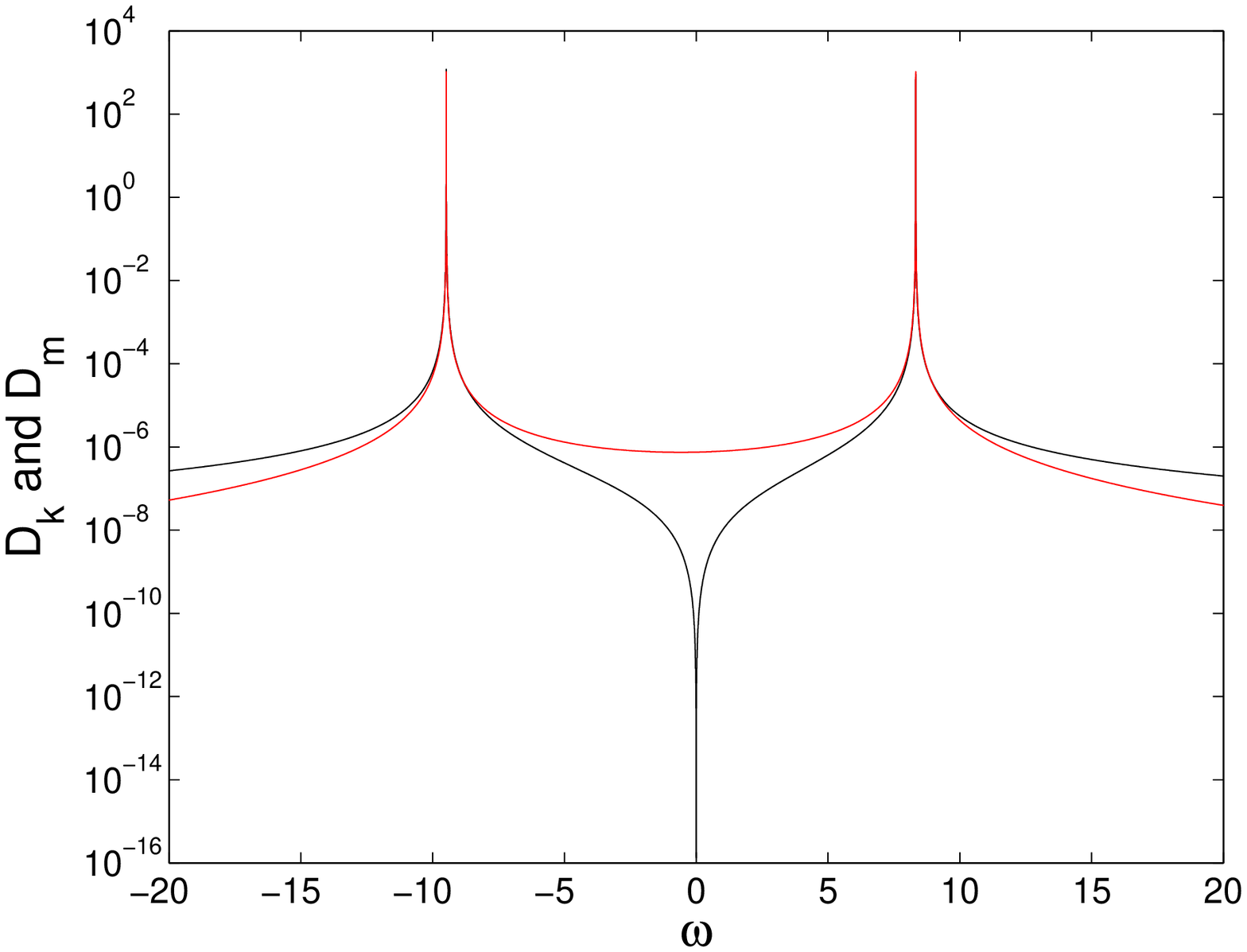}
\caption{The rotating MHD flow. The kinetic dissipation $D_k$ (black line) and the Ohmic dissipation $D_m$ (red line) versus the force frequency $\omega$. $E=10^{-6}$, $Le=1$, $Pm=1$, $\alpha=45^\circ$ and ($n_x=n_y=n_z=1$)}\label{fig:mhd1}
\end{figure}

After studying the rotating hydrodynamic flow, we move to the rotating MHD flow. Firstly we fix $E=10^{-6}$, $Le=1$, $Pm=1$, $\alpha=45^\circ$ and ($n_x=n_y=n_z=1$) to study the dependence of resonance on frequency. The two resonant frequencies are then $8.3272$ and $-9.4819$ (see equation \eqref{eq:resonant-frequency-solutions-dimensionless}), both of which are out of the range of inertial waves $(-2\le\omega\le2)$. According to equation \eqref{eq:resonant-frequency-solutions-dimensionless} the two resonant frequencies are almost proportional to $Le$ for high $Le$. Since the two resonant frequencies at $Le=1$ are already out of the range of inertial waves, $Le=1$ is high enough for the rotating MHD flow to be different from the rotating hydrodynamic flow. Figure \ref{fig:mhd1} shows the kinetic and Ohmic dissipations versus the force frequency. We can clearly see that the dissipations reach their peaks at the two resonant frequencies and their minima at $\omega=0$. To understand the difference between the kinetic and Ohmic dissipations at $\omega=0$, we come back to equations \eqref{eq:kinetic-dissipation-dimensionless} and \eqref{eq:ohmic-dissipation-dimensionless}. The two expressions differ by a pre-factor. When $\omega=0$, the denominator of the expression of $D_m$ \eqref{eq:ohmic-dissipation-dimensionless} is very small because of the small Ekman number. Therefore, even if $D_k$ is almost zero at $\omega=0$, $D_m$ is finite because of the very small denominator. Physically, it implies that a very low tidal frequency cannot lead to viscous dissipation but a certain amount of Ohmic dissipation. Moreover, as shown in table \ref{tab:mhd1}, at the positive resonant frequency $D_m$ is higher than $D_k$ whereas at the negative resonant frequency $D_k$ is higher than $D_m$, but $D_m$ keeps the same at both the positive and negative resonant frequencies. So we can sort the four dissipations as $D_k^->D_m^+=D_m^->D_k^+$, which we will see later in this section.

\begin{table}
\centering
\begin{tabular}{lll}
\hline
$\omega$ & $8.3272$ & $-9.4819$ \\
$D_k$    & $0.9230\times10^3$ & $1.1967\times10^3$ \\
$D_m$    & $1.0510\times10^3$ & $1.0510\times10^3$ \\
\hline
\end{tabular}
\caption{The rotating MHD flow. $D_k$ and $D_m$ versus the two resonant frequencies. The other parameters are the same as in figure \ref{fig:mhd1}.}\label{tab:mhd1}
\end{table}

\begin{figure}
\centering
\includegraphics[scale=0.4]{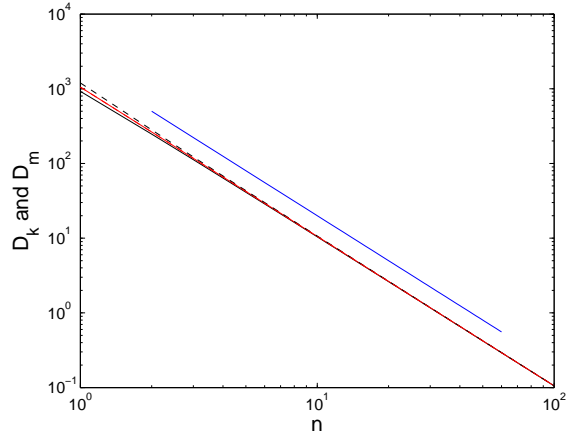}
\caption{The rotating MHD flow. The kinetic dissipation $D_k$ (black lines) and the Ohmic dissipation $D_m$ (red lines) versus the force wavenumbers $n_x=n_y=n_z=n$ at the two resonant frequencies. The solid lines denote the positive resonant frequency and the dashed lines denote the negative resonant frequency. The blue straight line shows the scaling $n^{-2}$. $E=10^{-6}$, $Le=1$, $Pm=1$ and $\alpha=45^\circ$.}\label{fig:mhd2}
\end{figure}

We next study the effect of the force wavenumber on the dissipations at the resonant frequencies. We keep the parameters the same as in the above study about the force frequency, i.e. $E=10^{-6}$, $Le=1$, $Pm=1$ and $\alpha=45^\circ$. As in the study of the rotating hydrodynamic flow, we keep the orientation of wavevector, i.e. $n_x=n_y=n_z$, but increase its amplitude. However, the MHD case is different from the hydrodynamic case in which the resonant frequency is determined merely by the orientation of wavevector. Now in the rotating MHD flow the resonant frequencies depend on both orientation and magnitude of wavevector (see equation \eqref{eq:resonant-frequency-solutions-dimensionless}). So we need to calculate firstly the two resonant frequencies at the given wavenumbers and then the dissipations at the two resonant frequencies. Figure \ref{fig:mhd2} shows the two dissipations versus the wavenumbers at the corresponding resonant frequencies. We can see that at the low wavenumbers $D_k^->D_m^+=D_m^->D_k^+$ but at the high wavenumbers all the four lines overlap, namely $D_k$ and $D_m$ at both the positive and negative resonant frequencies are equal on the small scales (in this case at the wavenumbers higher than $20$). Moreover, all the four dissipations scale as $k^{-2}$, which obeys the same scaling law of the rotating hydrodynamic flow. Again, in the rotating MHD flow, the resonance takes its effect on the dissipation at intermediate length scales.

\begin{figure}
\centering
\includegraphics[scale=0.4]{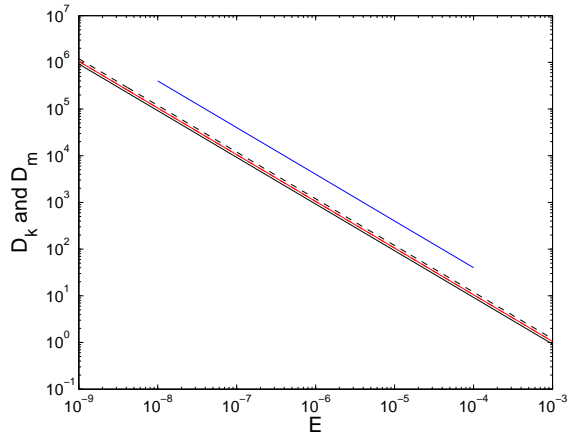}
\caption{The rotating MHD flow. The kinetic dissipation $D_k$ (black lines) and the Ohmic dissipation $D_m$ (red lines) versus the Ekman number $E$ at the two resonant frequencies. The solid lines denote the positive resonant frequency and the dashed lines denote the negative resonant frequency. The blue straight line shows the scaling $E^{-1}$. $Le=1$, $Pm=1$ and $\alpha=45^\circ$. $n_x=n_y=n_z=1$.}\label{fig:mhd3}
\end{figure}

We now study the resonance at the different dimensionless parameters $E$, $Le$ and $Pm$, and the angle $\alpha$. Firstly we fix $Le=1$, $Pm=1$ and $\alpha=45^\circ$ to study $E$. Because we have already known that the resonance is significant on the relatively large length scales, we fix the wavenumbers to be the lowest, i.e. $n_x=n_y=n_z=1$. Figure \ref{fig:mhd3} shows $D_k$ and $D_m$ versus $E$ at the two resonant frequencies. It shows that both $D_k$ and $D_m$ at both the positive and negative resonant frequencies scale as $E^{-1}$, which obeys the same scaling law of rotating hydrodynamic flow. It also shows that $D_k^->D_m^+=D_m^->D_k^+$. In summary, in rotating MHD flow, both $D_k$ and $D_m$ at the resonant frequencies scale as
\begin{equation}
D_k, D_m\propto E^{-1}k^{-2}.
\end{equation}

\begin{figure}
\centering
\includegraphics[scale=0.4]{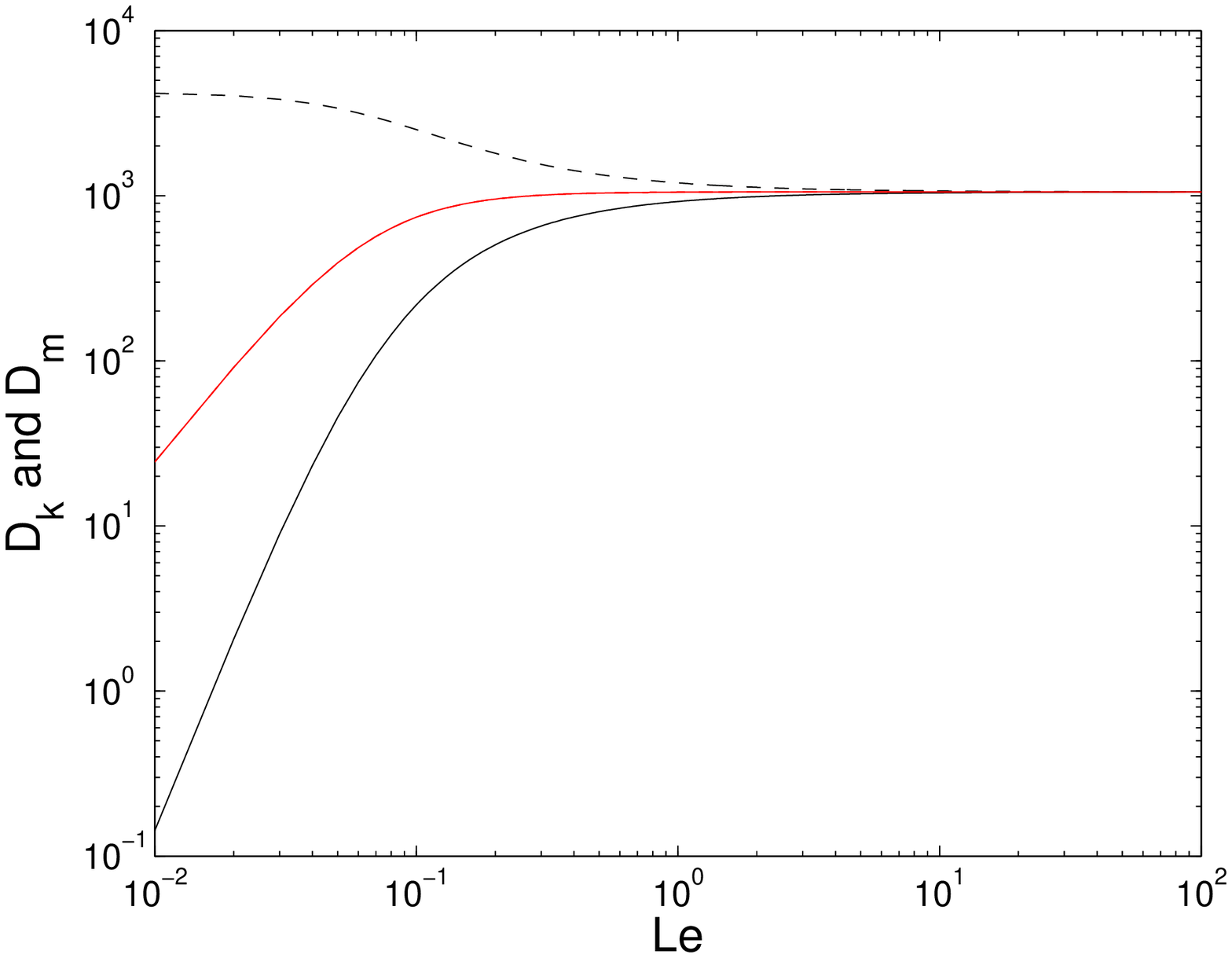}
\caption{The rotating MHD flow. The kinetic dissipation $D_k$ (black lines) and the Ohmic dissipation $D_m$ (red lines) versus the Lehnert number $Le$ at the two resonant frequencies. The solid lines denote the positive resonant frequency and the dashed lines denote the negative resonant frequency. $E=10^{-6}$, $Pm=1$ and $\alpha=45^\circ$. $n_x=n_y=n_z=1$.}\label{fig:mhd4}
\end{figure}

Next we keep $E$, $Pm$, $\alpha$ and the force wavenumbers to study $Le$. We increase $Le$ from $10^{-2}$ to $10^2$. Because the resonant frequencies depend on $Le$ (see equation \eqref{eq:resonant-frequency-solutions-dimensionless}), we need to firstly calculate the two resonant frequencies for the different Lehnert numbers and then the dissipations at the two resonant frequencies. Figure \ref{fig:mhd4} shows $D_k$ and $D_m$ versus $Le$ at the two resonant frequencies with the corresponding $Le$. As before, $D_m^+$ and $D_m^-$ are equal. At the high $Le$ ($>10$) all the four dissipations are almost equal. But at the low $Le$ ($<10^{-1}$), the kinetic dissipation at the negative resonant frequency $D_k^-$ is dominant, the kinetic dissipation at the positive resonant frequency $D_k^+$ is negligible, and the Ohmic dissipation $D_m^\pm$ is in between, i.e. $D_k^- > D_m^\pm > D_k^+$. Moreover, at the lower $Le$ this asymmetry is more striking. In the rotating hydrodynamic flow there is only one resonant frequency ($\omega_0=-2k_z/k$) but in the rotating MHD flow the dissipations are asymmetric to the positive and negative resonant frequencies $\omega^{\pm}$. Therefore, this asymmetry apparently arises from the magnetic field.

\begin{figure}
\centering
\includegraphics[scale=0.4]{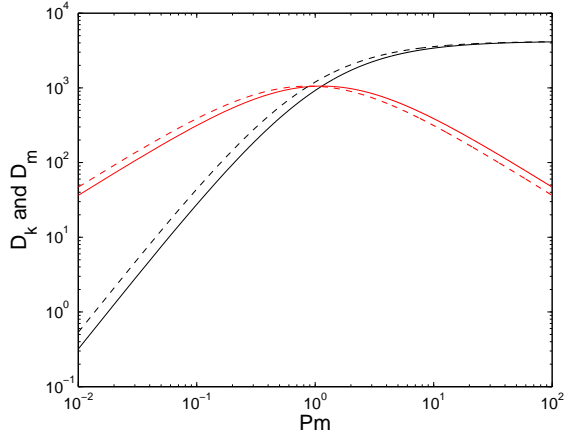}
\caption{The rotating MHD flow. The kinetic dissipation $D_k$ (black lines) and the Ohmic dissipation $D_m$ (red lines) versus the magnetic Prandtl number $Pm$ at the two resonant frequencies. The solid lines denote the positive resonant frequency and the dashed lines denote the negative resonant frequency. $E=10^{-6}$, $Le=1$ and $\alpha=45^\circ$. $n_x=n_y=n_z=1$.}\label{fig:mhd5}
\end{figure}

We then study the magnetic Prandtl number $Pm=\nu/\eta$ which measures the relative strength of the two dissipation mechanisms. As usual, we keep $E$, $Le$, $\alpha$ and the force wavenumbers but increase $Pm$ from $10^{-2}$ to $10^2$. Figure \ref{fig:mhd5} shows $D_k$ and $D_m$ versus $Pm$ at the two resonant frequencies. It is not surprising that the viscous (or Ohmic) dissipation is higher than Ohmic (or viscous) dissipation at $Pm>1$ (or $Pm<1$) and the two dissipations are close to each other at $Pm=1$. However, it is interesting that with $Pm$ increasing $D_k$ increases monotonically whereas $D_m$ increases until $Pm=1$ and then decreases. Moreover, $D_m^+>D_m^-$ at $Pm>1$, $D_m^->D_m^+$ at $Pm<1$, and $D_m^+=D_m^-$ at $Pm=1$. In geophysical and astrophysical MHD flows, $Pm<1$, and so Ohmic dissipation is more important than viscous dissipation.

\begin{figure}
\centering
\includegraphics[scale=0.4]{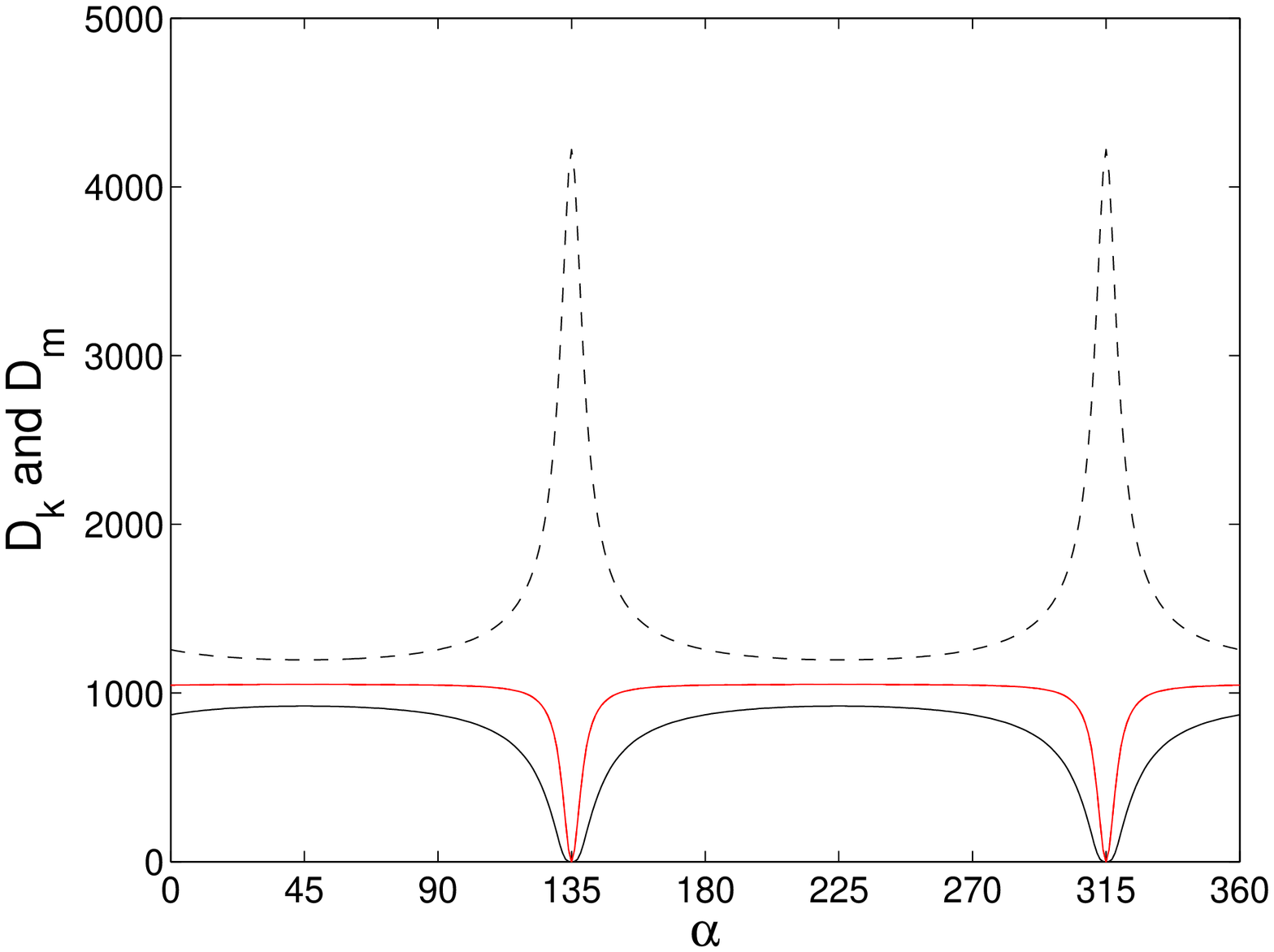}
\caption{The rotating MHD flow. The kinetic dissipation $D_k$ (black lines) and the Ohmic dissipation $D_m$ (red lines) versus the angle $\alpha$ at the two resonant frequencies. The solid lines denote the positive resonant frequency and the dashed lines denote the negative resonant frequency. $E=10^{-6}$, $Le=1$ and $Pm=1$. $n_x=n_y=n_z=1$.}\label{fig:mhd6}
\end{figure}

We next study the angle $\alpha$ between the rotation and the magnetic field. Figure \ref{fig:mhd6} shows $D_k$ and $D_m$ versus $\alpha$ with the other parameters fixed. At $\alpha=135^\circ$ and $315^\circ$, $D_k^-$ at the negative resonant frequency dominates while the other three dissipations are negligible. These two angles for the maximum and minimum of the dissipations are determined by the orientation of the wavevector, i.e. the factor ($k_x\sin\alpha+k_z\cos\alpha$) in the formulae to calculate dissipations. The two angles can be deduced to be
\begin{equation}
\pi-\arctan(k_z/k_x) \hspace{3mm} \mbox{and} \hspace{3mm} 2\pi-\arctan(k_z/k_x),
\end{equation}
which indicates that the wavevector is perpendicular to the magnetic field. Moreover, when $\bm k$ is perpendicular to $\bm B_0$ the Alfv\'en frequency $\omega_B$ will be zero such that the resonant frequency of the rotating MHD flow is equal to that of the rotating hydrodynamic flow, see \eqref{eq:resonant-frequency} and \eqref{eq:resonant-frequency-rotating}, and then the dissipation of rotating MHD flow is also equal to that of the rotating hydrodynamic flow. It should be noted that this result is valid only at the resonant frequencies.

\begin{figure}
\centering
\includegraphics[scale=0.65]{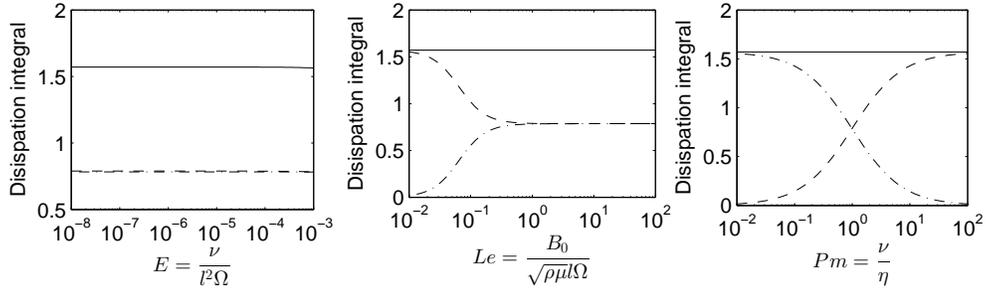}
\caption{The rotating MHD flow. The dissipation integrals over frequency over Ekman number, Lenert number and magnetic Prandtl number. The solid lines denote the total dissipation, the dashed lines the viscous dissipation and the dash-dotted lines the Ohmic dissipation.}\label{fig:integral}
\end{figure}

After the investigation of dissipation at particular frequencies, we study the dissipation integral over frequency. In the rotating hydrodynamic flow, by virtue of \eqref{eq:resonance-dimensionless}, \eqref{eq:sigma-dimensionless} and \eqref{eq:kinetic-dissipation-dimensionless}, we can readily derive the viscous dissipation integral
\begin{equation}
\int_{-\infty}^{\infty}D_kd\omega=\int_{-\infty}^{\infty}\frac{E}{2}\frac{|ik\hat{\bm f}|^2}{|\omega-\omega_0+iEk^2|^2}d\omega=\frac{E}{2}k^2a^2\frac{1}{Ek^2}\int_{-\infty}^{\infty}\frac{d\left(\frac{\omega-\omega_0}{Ek^2}\right)}{1+\left(\frac{\omega-\omega_0}{Ek^2}\right)^2}=\frac{\pi}{2}a^2.
\end{equation}
The numerical integration with the accurate Gauss-Legendre method yields the same result. In the rotating MHD flow, the analytical derivation is not straightforward and we do the numerical integration. Figure \ref{fig:integral} shows the integrals of viscous, Ohmic and total dissipations over a wide range of frequency (from -4000 to 4000). It verifies that the total dissipation is always $\pi/2$ and independent of viscosity (Ekman number), imposed magnetic field (Lenert number) and magnetic diffusivity (magnetic Prandtl number). The left panel shows that the viscous and Ohmic dissipations are independent of Ekman number (both are $\pi/4$ at $Le=1$ and $Pm=1$) and the total dissipation is exactly $\pi/2$. The middle panel shows that at small $Le$ the viscous dissipation integral dominates over the Ohmic dissipation integral whereas at large $Le$ ($>1$) the two dissipation integrals reach the same level $\pi/4$. It is not surprising that the Ohmic dissipation integral dominates over the viscous dissipation integral for $Pm<1$ and vice versa for $Pm>1$, as shown in the right panel. The fact that the total dissipation integral is constant can be interpreted as follows. Consider a damped harmonic oscillator $\ddot{x}+\gamma\dot{x}+\omega_0^2x=a\cos(\omega t)$ where $\gamma$ is the friction coefficient, $\omega_0$ is the natural frequency and $\omega$ is the forcing frequency. We can use this toy model to understand the tidal dissipation in the fluid system. $\gamma$ is analogy to viscosity and magnetic diffusivity, $\omega_0$ is the eigen-frequency of inertial or magneto-inertial wave, and $\omega$ is the tidal frequency. By using Green's function we can find the dissipation rate of this damped harmonic oscillator to be $D=2a^2\gamma\omega^2/[(\omega^2-\omega_0^2)^2+(\gamma\omega)^2]$. The integral of dissipation rate over all forcing frequencies $\int_{-\infty}^\infty Dd\omega=2\pi a^2$ is thus constant depending only upon the forcing amplitude $a$ but independent of the friction coefficient $\gamma$ and the natural frequency $\omega_0$. That is, the frequency-averaged dissipation is constant, and this result may have important astrophysical consequences.

\section{Applications to astrophysics}\label{sec:application}

We can apply these results to the tidal dissipation in the geophysical and astrophysical fluids. Firstly, in the presence of magnetic field, the range of the resonant frequencies is broader, i.e. $\omega^-<\omega_0<\omega^+$, and out of $-2\Omega\le\omega\le2\Omega$ in the purely hydrodynamic flow (for example, with the parameters in this paper it changes from $-1.1547$ to $8.3272$ and $-9.4819$). Therefore, the tidal resonance is more likely to occur. Secondly, the dissipation at the resonance on the small length scales is insignificant and equal to the dissipation of the non-resonance. So the major contribution to the tidal dissipation at the resonance is at intermediate length scales rather than small scales. Usually the dissipation on the small scales is stronger than on the large scales, but at the resonant frequency the situation is opposite, namely dissipation scales as $E^{-1}k^{-2}$. Thirdly, the more rapid rotation or the smaller viscosity (the lower $E$) leads to the higher dissipation at the resonance. That the smaller viscosity leads to the higher dissipation is opposite to the intuition, which is again because of the resonance. Fourthly, the dissipation at the negative resonant frequency dominates over the other dissipations for the small $Le$, and this might happen in the rapidly rotating and weakly magnetized stars. Fifthly, when the phase velocity of the inertial and magneto-inertial waves in the rapidly rotating fluid is perpendicular to the magnetic field, the amplitudes of the waves at the resonant frequencies will reach their maximum and hence the waves at the resonance will be highly damped.

Furthermore, we estimate the parameters $E$, $Le$ and $Pm$ used in our calculations. In the interiors of stars and giant planets, magnetic diffusivity is much larger than viscosity and $Pm$ is very small. Take the Sun and Jupiter for example. In solar convective zone $Pm$ is of order of $10^{-6}$ and in Jupiter's interior it is of the order of $10^{-4}$. Therefore, Ohmic dissipation is much stronger than viscous dissipation. By inserting the radius, rotation rate and viscosity of the Sun and Jupiter to the definition of Ekman number $E$, we estimate $E$ of the Sun and Jupiter to be both of the order of $10^{-16}$ which is very small (if we take $l$, the box size, to be $1/10$ or $1/100$ of the radius, $E$ is still very small). The small Ekman number implies the strong dissipation at resonance (remember $D_{k,m}\propto E^{-1}$). The Lehnert number is not easy to be estimated because the magnetic field in the interiors of the Sun and Jupiter is unknown. Although we know the surface field strength, the field in the interiors may be much stronger than surface because of the dynamo action, e.g. the strong differential rotation shears the poloidal field to create a strong toroidal field. 

\begin{figure}
\centering
\includegraphics[scale=0.4]{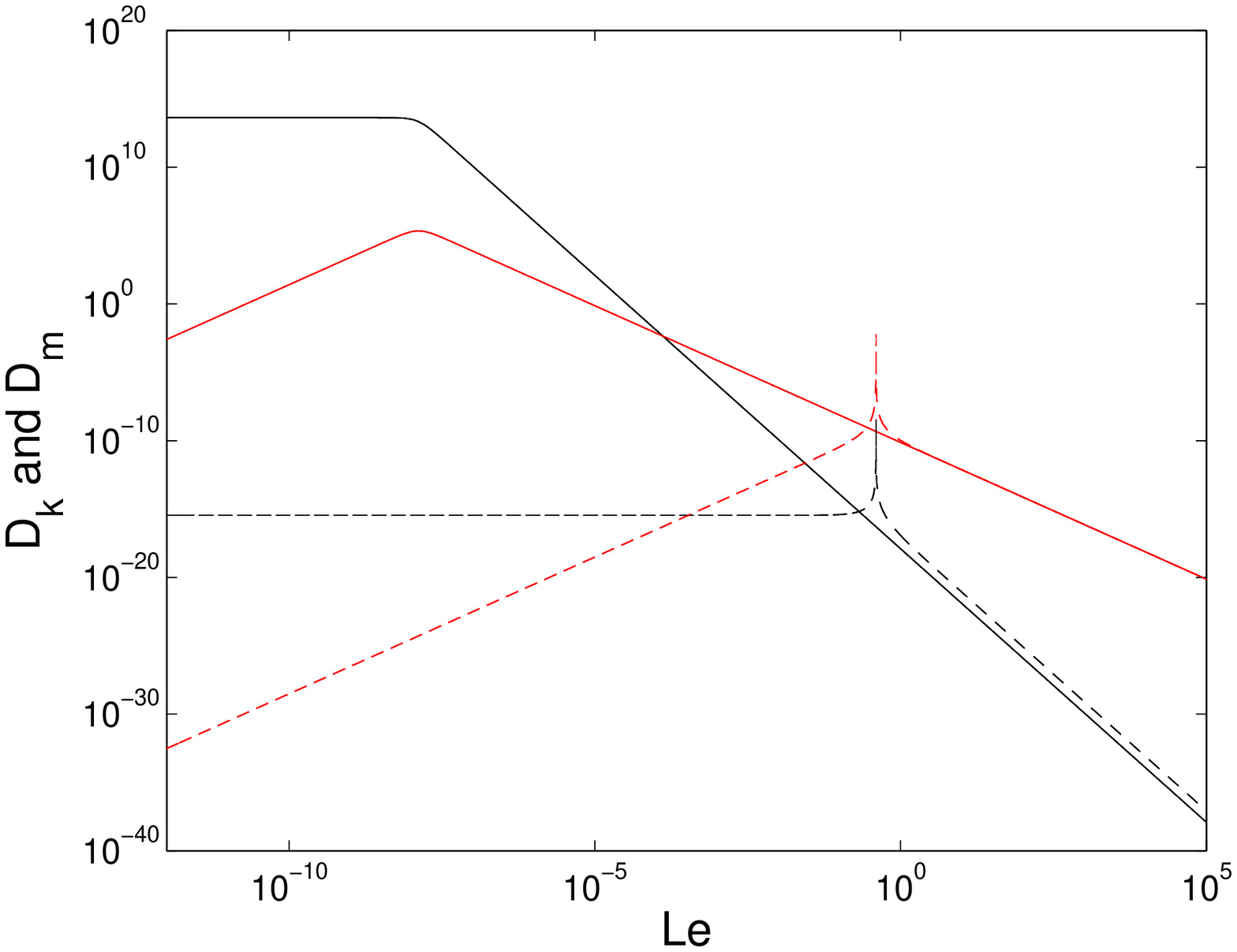}
\caption{The rotating MHD flow. The kinetic dissipation $D_k$ (black line) and the Ohmic dissipation $D_m$ (red line) versus $Le$. Solid lines denote $\omega=-2k_z/k=-2/\sqrt{3}$ and dashed lines $\omega=3.0$. $E=10^{-16}$, $Pm=10^{-6}$ and $\alpha=45^\circ$. $n_x=n_y=n_z=1$.}\label{fig:application}
\end{figure}

To understand the magnetic effect on tidal dissipation in stars and giant planets we scan the Lehnert number at a fixed frequency. We keep $E=10^{-16}$ and $Pm=10^{-6}$, both of which are as small as in the Sun or Jupiter. We choose two force frequencies to calculate dissipations. One frequency is chosen to be the resonant frequency of hydrodynamic flow, $\omega=\omega_0=-2k_z/k$. We choose this frequency in order to compare with the purely hydrodynamic flow. The other frequency is chosen to be $\omega=3.0$ at which the inertial wave cannot be excited by the tidal force such that only the magneto-inertial wave can be excited. As before, we take the wavenumbers to be $n_x=n_y=n_z=1$ and $\alpha=45^\circ$. The dissipation of purely hydrodynamic flow ($Le=0$) can be calculated to be $D_k=4.22\times10^{13}$ at $\omega_0$ and $D_k=3.43\times10^{-16}$ at $\omega=3.0$. Figure \ref{fig:application} shows the kinetic and Ohmic dissipations versus $Le$ at the given parameters which are close to those of the Sun and Jupiter. $Le$ is taken to be from $10^{-12}$ to $10^5$, a large range which covers the field strength on the surface and in the interior. The figure shows that at small $Le$ in the weak field regime ($Le\lesssim10^{-8}$ for $\omega_0$ or $Le\lesssim10^{-5}$ for $\omega=3.0$) the kinetic dissipations at the both frequencies $\omega_0$ and $\omega=3.0$ are equal to the ones of purely hydrodynamic flow whereas the Ohmic dissipations are negligible compared to the kinetic dissipations. However, at large $Le$ in the strong field regime ($Le\gtrsim1$) both the kinetic and Ohmic dissipations decrease with increasing $Le$ (the spikes for $\omega=3.0$ arise from resonances at $Le\approx0.4$), and the kinetic dissipations decay as $Le^{-4}$ and the Ohmic dissipations as $Le^{-2}$. These two scalings can be readily obtained by \eqref{eq:kinetic-dissipation-dimensionless}, \eqref{eq:ohmic-dissipation-dimensionless}, \eqref{eq:resonance-dimensionless} and \eqref{eq:sigma-dimensionless} under the condition $Le\gg1$. Then it is more interesting to study the intermediate range of $Le$. When $Le$ is of the order of $10^{-3}$ the Ohmic dissipations win out the kinetic dissipations at the both frequencies. We may then have a tentative result: \emph{in the regions of stellar and planetary interiors where the order of $Le$ is larger than $10^{-3}$, the magnetic effect on tidal dissipation should be considered}. In white dwarfs and neutron stars the magnetic fields are very strong, e.g. the surface fields of white dwarf can exceed $10^6$ Gauss, and therefore, in these compact objects it is very likely that the Ohmic dissipation dominates over the kinetic dissipation. Thus \emph{in the binary compact objects the magnetic effect on tidal dissipation should be considered}.

\section{Discussions and conclusions}\label{sec:discussion}

In this work we derived the linear response to the tidal forcing in the rotating MHD flow under the local WKB approximation, and then calculated the kinetic dissipation in the rotating hydrodynamic flow as well as both the kinetic and Ohmic dissipations in the rotating MHD flow. We focused on resonances and studied one by one the effects of the frequency, the wavenumber and the other parameters, namely the Ekman number, the Lehnert number, the magnetic Prandtl number and the angle $\alpha$. In the rotating hydrodynamic flow there is only one resonant frequency and the kinetic dissipation at the resonant frequency scales as $E^{-1}k^{-2}$. In the rotating MHD flow there exist two resonant frequencies, one positive and the other negative. In both the rotating hydrodynamic and MHD flows, the resonance takes its effect on the dissipation at intermediate length scales. In the rotating MHD flow, in the weak field regime (in terms of $Le<1$) the kinetic dissipation at the negative resonant frequency dominates over the other three dissipations ($D_k^- > D_m^\pm > D_k^+$), and all the four dissipations at the resonant frequencies scale as $E^{-1}k^{-2}$, which is the same as in the rotating hydrodynamic flow. The Ohmic dissipation exceeds the viscous dissipation at $Pm<1$ whereas the viscous dissipation exceeds the Ohmic dissipation at $Pm>1$. The wave damping at the resonance reaches its maximum when the wavevector is perpendicular to the magnetic field. In addition, we also find that the frequency-integrated total dissipation is constant, and that Ohmic dissipation is important for $Le>10^{-3}$.

It should be noted that studies of the magneto-elliptic instability due to tides suggest that a strong field increases dissipation and reduces synchronization time, see \citet{mizerski-bajer2011}. However, in our study a strong field decreases the ``diffusion'' as shown by figure \ref{fig:mhd4}, namely the dissipation at small $Le$ never gets close to the hydrodynamic value. This difference could arise from three possibilites. The first is the different dissipation mechanisms. In \citet{mizerski-bajer2011} both the molecular viscosity $\nu$ and the magnetic diffusivity $\eta$ are absent, the dissipation arises from the turbulent viscosity (i.e. turbulent Reynolds stress), and the enhancement of the dissipation is because of the addition of turbulent Maxwell stress. However, in our study the dissipation arises from the molecular viscosity and the magnetic diffusivity but not the turbulent stresses. The second is that in \citet{mizerski-bajer2011} the tidal dissipation is calculated with the elliptical instability whereas in our study it is directly calculated from the flow driven by the time-dependent dynamical tide ($\bm f$). The third is the elliptical effect. In \citet{mizerski-bajer2011} the elliptical streamline is assumed. However, in our study we do not consider the elliptical instability.

One may argue that the unbounded geometry is too simple. In a domain with boundaries, the inertial waves reflect and the internal thin layers form in which the dissipation is very strong. However, the addition of a magnetic field revises the Poincar\'e equation governing the inertial waves in a rotating fluid, such that the waves will not focus in the internal thin shear layers, see \citep{tilgner2000}. Secondly, we need to clarify again that this work only begins the investigation of magnetic effects on dynamical tides and we will carry out more work in spherical geometry.

\section*{Acknowledgements}
This work was initiated in Princeton and completed in Shanghai. Prof. Jeremy Goodman gave me valuable suggestions. The anonymous referee gave me valuable suggestions and comments, and revised the manuscript and corrected the grammar mistakes in details. I am financially supported by the National Science Foundation’s Center for Magnetic Self-Organization under grant PHY-0821899 and the startup grant WF220441903 of Shanghai Jiao Tong University.

\bibliographystyle{apj}
\bibliography{paper}

\end{document}